# Random Forest for Malware Classification

Felan Carlo C. Garcia
Department of Electronics, Computer
and Communications Engineering
Ateneo de Manila University
Quezon City, Philippines
felan.garcia@obf.ateneo.edu

Felix P. Muga II, PhD.
Mathematics Department
Ateneo de Manila University
Quezon City, Philippines
fmuga@ateneo.edu

*Abstract* — *The challenge in engaging malware activities involves the correct identification and classification of different malware variants. Various malwares incorporate code obfuscation methods that alters their code signatures effectively countering anti-malware detection techniques utilizing static methods and signature database. In this study, we utilized an approach of converting a malware binary into an image and use Random Forest to classify various malware families. The resulting accuracy of 0.9562 exhibits the effectivess of the method in detecting malware.*

## I. Introduction

Cybercrime operations through networked computer systems remains a growing threat for developed regions with a mature information and communications (ICT) infrastructure in which a considerable number of public and private services are dependent. At the core of Cybercrime operations are Malwares consisting of spywares, bots, rootkits, Trojan, and viruses designed to perform tasks such as service disruption, network hijacking, exploiting resources, and private information stealing [1].

The challenge in engaging malware activities involves the correct identification and classification of different malware variants. Malwares incorporate code obfuscation and metamorphism to change their code signatures while maintaining their behaviors and functionalities [2]. These methods effectively counters anti-malware software relying on malware signature database to identify a specific malware attacking a computer system. Lastly, this code obfuscation and morphing generates a high volume of data points for a certain malware variant alone [3].

Nataraj et al [4] proposed a method on visualizing malware binaries as image file resulting on complex visual patterns acting as a malwares signature. The obfuscation of the code also introduces various changes on the resulting image but still retain the general structure and thus show potential as an approach to classifying malware variants.

In this study, we take advantage of malware as image files as feature vectors and Random Forest to effectively classify and segregate malware families from each other.

## II. Dataset

In this study we evaluate our methods on the Malimg Dataset [4] consisting of 9,342 malware samples of 25 different malware families. Table 1 shows the malware families consisting the Malimg Dataset and the equivalent population % of each of the families within the data set. It is worth noting that the following dataset is imbalanced and developing a training set must include a stratified sampling of the populations to prevent overfitting and under generalization on specific malware variants.

| Table 1: Malimg Dataset | | | | |
|---|---|---|---|---|
| No. | Family | Family Name | No. of Variants | Population % |
| 1 | Worm | Allaple.L | 1591 | 0.17 |
| 2 | Worm | Allaple.A | 2949 | 0.316 |
| 3 | Worm | Yuner.A | 800 | 0.086 |
| 4 | PWS | Lolyda.AA 1 | 213 | 0.023 |
| 5 | PWS | Lolyda.AA 2 | 184 | 0.02 |
| 6 | PWS | Lolyda.AA 3 | 123 | 0.013 |
| 7 | Trojan | C2Lop.P | 146 | 0.016 |
| 8 | Trojan | C2Lop.gen!G | 200 | 0.021 |
| 9 | Dialer | Instantaccess | 431 | 0.046 |
| 10 | Trojan Downloader | Swizzor.gen!I | 132 | 0.014 |
| 11 | Trojan Downloader | Swizzor.gen!E | 128 | 0.014 |
| 12 | Worm | VB.AT | 408 | 0.044 |
| 13 | Rogue | Fakerean | 381 | 0.041 |
| 14 | Trojan | Alueron.gen!J | 198 | 0.021 |
| 15 | Trojan | Malex.gen!J | 136 | 0.015 |
| 16 | PWS | Lolyda.AT | 159 | 0.017 |
| 17 | Dialer | Adialer.C | 125 | 0.013 |
| 18 | Trojan Downloader | Wintrim.BX | 97 | 0.01 |
| 19 | Dialer | Dialplatform.B | 177 | 0.019 |
| 20 | Trojan Downloader | Dontovo.A | 162 | 0.017 |
| 21 | Trojan Downloader | Obfuscator.AD | 142 | 0.015 |
| 22 | Backdoor | Agent.FYI | 116 | 0.012 |
| 23 | Worm:AutoIT | Autorun.K | 106 | 0.011 |
| 24 | Backdoor | Rbot!gen | 158 | 0.017 |
| 25 | Trojan | Skintrim.N | 80 | 0.009 |

Table I: Malware families comprising the Malimg Dataset.

## III. METHODOLOGY

### A. Data Preparation

Malware binaries sequences are grouped as 8-bit vectors. The resulting 8-bit vectors are then plotted as a grayscale image as shown on Fig 1.

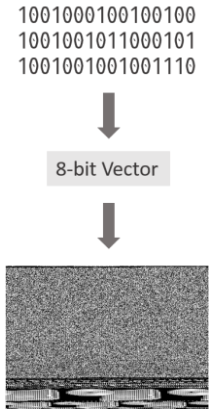

Figure 1: Conversion of Malware binaries into gray scale images.

The conversion result of the malware binaries consists of images with different sizes and patterns as shown on Fig 2. It is also worth noting that several malware creators also add extra binary code pattern to their malware as a personal signature such as the bottom part of the image shown in Fig 2.C

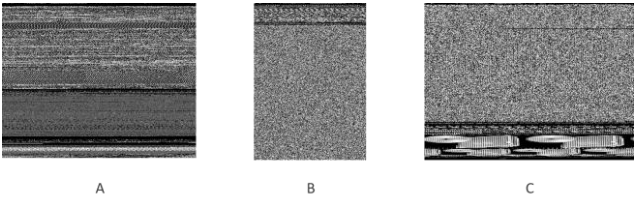

Figure 2: Resulting Malware images from the binary files. The following families above are: A) Adialer.C B) Malex.gen!J, C) Fakerean.

The malware images are resized into a 2-dimensional matrix to have a uniform dataset. The resized images are flattened into *n x n* array where *n* = 32. Each resulting array with length of 1024 is labelled with its corresponding malware family.

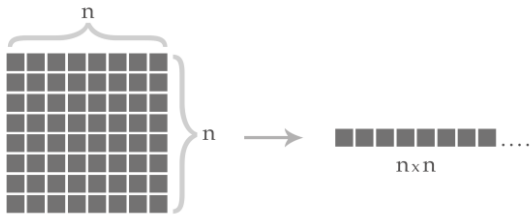

Figure 3: Conversion of an *n x n* image matrix into a flat array.

The labeled arrays are appended together as a row into a csv file that would comprise the training set for the machine learning algorithm.

### B. Classification

Various researches on malware and cyber anomaly detections utilized machine learning methods such as Support Vector Machines (SVM), K-Nearest Neighbors (K-NN), and Neural Networks (NN) [6], for this study, we utilized the use of Random Forest as a feasible method for malware classification.

In terms of supervised learning and performances various studies have ranked Gradient Boosted Trees, Random Forests, Neural Networks, and Support Vector Machines to have high predictive accuracies [7][8]. While Gradient Boosted Trees did have the highest accuracy, Random Forest was able to achieve almost the same performance with minor parameter tuning [7].

In this study, we utilized the Random Forest implementation on R with the randomForest and caret library.

### C. Training and Validation

Creating the training and testing set involves splitting the data into 80% training and 20% testing. The splitting of the dataset also involves taking into account the relative populations of each malware families to ensure that each family are well represented on the split dataset.

A k-fold Cross-Validation procedure is used to evaluate the model where the training data is randomly partition into different subsamples with equal k sizes. One k subsample is held out as validation data and the remaining k subsamples are used as training data. This process is then repeated k-times (referred as the number of folds) with each of the k subsample used as validation. The resulting accuracies for each fold is averaged to produce a single estimation of the models accuracy for a particular machine learning problem [9].

A 10-fold Cross-Validation procedure performed on the training set to evaluate the model, afterwards the model is tested on the held-out testing set and evaluated for its performance.

## IV. RESULTS

The training set for the data consist of a 1024 feature vector with a corresponding label. We first evaluate the cross-validation results of the model with the training set.

**Table 2: Cross Validation Result**

| Accuracy | Kappa | Accuracy Lower | Accuracy Upper |
|---|---|---|---|
| 0.9464 | 0.9367 | 0.9411 | 0.9514 |

Table 2: Summary of the 10-Fold Cross Validation accuracy metrics

The resulting metrics as shown on Table 2 indicates a strong predictive performance from the model. The model's overall predictive accuracy is *0.9464* within the bounds of [*0.9411, 0.9514*]. Another metric considered is the Kappa statistic which indicates if the proximity of the instances classified by the predictive model matched the testing data's ground truth [10].

The measured Kappa for the cross validation result is 0.9367 and provides a strong indication with regards to the accuracy of the random forest model for the training set. Fig 4. Illustrates the confusion matrix of the cross validation results. Majority of the malware families have accuracies in the range of 0.9 and above. It is worth noting that 4 malware families (CL2OP.gen!g, C2LOP.P, Swizzor.gen!E, Swizzor.gen!I) have accuracies that fall below 0.5.

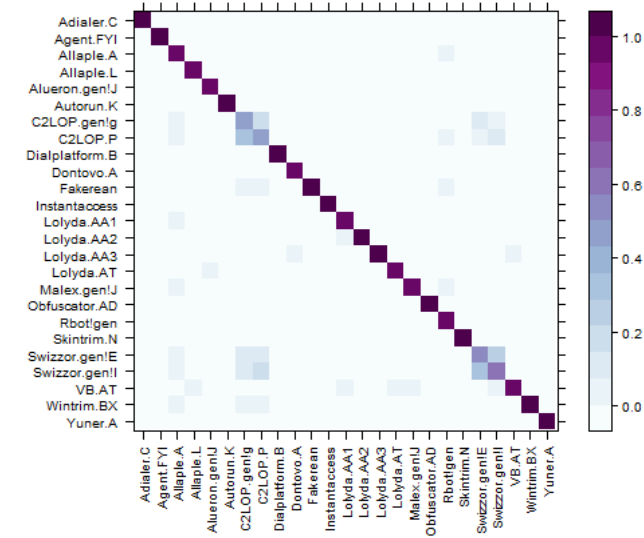

Figure 4: Cross Validation Confusion Matrix. The figure shows the predictive accuracy for each malware families with a score of 0.0 – 1.0 where 1.0 represents that all samples for the particular family are classified correctly.

The random forest model is then tested on the held-out testing set consisting of the remaining 20% of the data un-seen during the training phase of the model. Table 3 summarizes the resulting metrics of the model utilizing the testing set. The overall accuracy of the model is *0.9526* within the bounds within the bounds of [*0.9411, 0.9514*]. The measured Kappa is 0.9441 and indicates the predictive strength of the random forest model. The testing set results are similar to the results of the cross validation. The results exhibit the reliability of cross validation in measuring the predictive strength of a model prior to testing the model to an unseen data.

**Table 3: Held out Testing Set Result**

| Accuracy | Kappa | Accuracy Lower | Accuracy Upper |
|---|---|---|---|
| 0.9526 | 0.9441 | 0.9420 | 0.9618 |

Table 3: Summary of the testing set accuracy metrics

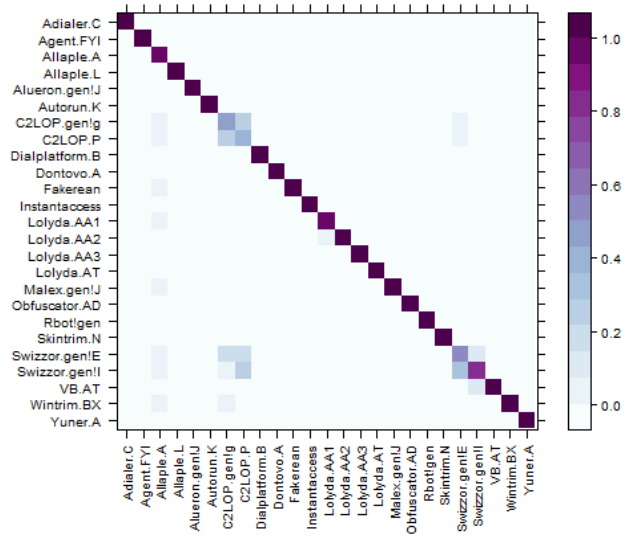

Figure 5: Held-out Testing Set Confusion Matrix. The figure shows the predictive accuracy for each malware families with a score of 0.0 – 1.0 where 1.0 represents that all samples for the particular family are classified correctly.

The confusion matrix as shown on Figure 5. Shows the resulting accuracies for each malware families and exhibits the same pattern particularly on the 4 malware families (CL2OP.gen!g, C2LOP.P, Swizzor.gen!E, Swizzor.gen!I) having less than 0.5 accuracy. Furthermore both Fig 4 and Fig 5 show that CL2OP.gen!g, C2LOP.P, Swizzor.gen!E, and Swizzor.gen!I malware families exhibit misclassifications between each other.

Inspecting the image data for each of the families reveals that these malware families exhibit similar visual patterns as shown on Fig 6. The visual similarity of the images coupled with the image resizing procedure would likely result into a training and testing set with similar data points for each of the malware families. This visual similarity would likely explain why the misclassifications were concentrated on the 4 families as shown on the confusion matrices in Fig 4 and Fig 5.

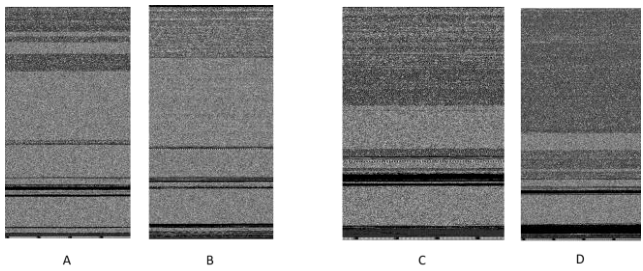

Figure 6: Visual inspection of the malware families of : A) CL2OP.gen!g, B) C2LOP.P, C) Swizzor.gen!E, and D)Swizzor.gen!I. exhibiting similar visual patterns

## V. CONCLUSION AND FUTURE WORK

In this study we exhibited the used of malware images as a feature vector for classifying various malware families. The study used Random Forest and performed 10-fold Cross Validation to determine the predictive strength of the model The resulting accuracies have shown that Random Forest model achieved a 0.9526 classification accuracy for the given malware dataset. However, it is also worth noting that there are still things to consider such as misclassification on visually similar malware families.

As recommendation, since the study has used the image from the malwares as the only features for the training set, future works on the topic can use feature extraction utilizing image processing which can provide additional insights and better training procedures for the model.